# A Fuzzy Approach for Feature Evaluation and Dimensionality Reduction to Improve the Quality of Web Usage Mining Results

Zahid Ansari[#1], M.F.Azeem[*2], A. Vinaya Babu[‡3] and Waseem Ahmed[#4]

[#] *Department of Computer Science, P.A. College of Engineering, Mangalore, India*
E-mail: [1]*zahid.ansari@acm.org*, [4]*waseem@computer.org*

[*]*Department of Electronics and Communication, P.A. College of Engineering, Mangalore, India*
E-mail: [2]*mf.azeem@gmail.com*

[‡]*Department of Computer Science, J.N.T. University, Hyderabad, India*
E-mail: [3]*dravinayababu@jntuh.ac.in*

*Abstract—* The explosive growth in the information available on the Web has necessitated the need for developing Web personalization systems that understand user preferences to dynamically serve customized content to individual users. Web server access logs contain substantial data about the accesses of users to a Web site. Hence, if properly exploited, the log data can reveal useful information about the navigational behaviour of users in a site. In order to reveal the information about user preferences from, Web Usage Mining is being performed. Web Usage Mining is the application of data mining techniques to web usage log repositories in order to discover the usage patterns that can be used to analyze the user's navigational behavior. WUM contains three main steps: preprocessing, knowledge extraction and results analysis. During the preprocessing stage, raw web log data is transformed into a set of user profiles. Each user profile captures a set of URLs representing a user session. Clustering can be applied to this sessionized data in order to capture similar interests and trends among users' navigational patterns. Since the sessionized data may contain thousands of user sessions and each user session may consist of hundreds of URL accesses, dimensionality reduction is achieved by eliminating the low support URLs. Very small sessions are also removed in order to filter out the noise from the data. But direct elimination of low support URLs and small sized sessions may results in loss of a significant amount of information especially when the count of low support URLs and small sessions is large. We propose a fuzzy solution to deal with this problem by assigning weights to URLs and user sessions based on a fuzzy membership function. After assigning the weights we apply a "Fuzzy c-Mean Clustering" algorithm to discover the clusters of user profiles. In this paper, we describe our fuzzy set theoretic approach to perform feature selection (or dimensionality reduction) and session weight assignment. Finally we compare our soft computing based approach of dimensionality reduction with the traditional approach of direct elimination of small sessions and low support count URLs. Our results show that fuzzy feature evaluation and dimensionality reduction results in better performance and validity indices for the discovered clusters.

## I. INTRODUCTION

The World Wide Web as a large and dynamic information source is a fertile ground for data mining principles or Web Mining. Web mining is primarily aimed at deriving actionable knowledge from the Web through the application of various data mining techniques [1]. Web Usage Mining is the discovery of user access patterns from Web server access logs [2]. Web Usage Mining analyses results of user interactions with a Web server, including Web logs and database transactions at a Web site. Web usage mining includes clustering to find natural groupings of users or pages, associations to discover the URLs requested together and analysis of the sequential order in which URLs are accessed [3]. Web Usage Mining (WUM) consists of three main steps: preprocessing, knowledge extraction and results analysis [4]. The goal of the preprocessing step is to transform the raw web log data into a set of user profiles. Each such profile captures a sequence or a set of URLs representing a user session [5]. Web usage data preprocessing exploit a variety of algorithms and heuristic techniques for various preprocessing tasks such as data cleaning, user identification, session identification etc [6]. Data cleaning involves tasks such as, removing extraneous references to embedded objects, style files, graphics, or sound files, and removing references due to spider navigations [7]. User identification refers to the process of identifying unique users from the user activity logs [8]. User Session identification is the process of segmenting the user activity log of each user into sessions, each representing a single visit to the site [9]. Once user sessions are discovered, this sessionized data can be used as the input for a variety of data mining tasks such as clustering, association rule mining, sequence mining etc. If the data mining task at hand is clustering, the session files are filtered to remove very small sessions in order to eliminate the noise from the data [10]. But direct removal of these small sized sessions may result in loss of a significant amount of information especially when the number of small sessions is large. We propose a "Fuzzy Set Theoretic" approach to deal with this problem. Instead of directly removing all the small sessions below a specified threshold, we assign weights to all the sessions using a "Fuzzy Membership Function" based on the number of URLs accessed by the sessions. Moreover the URLs appearing in the access logs could number in the thousands. Clustering methods often perform very poor when dealing with very high dimensional data. Therefore filtering the logs by removing the URLs that are not supported by sufficient number of user sessions can provide an effective dimensionality reduction method while improving clustering results. We propose a "Fuzzy Set Theoretic" approach for filtering the low support URLs. We assign weights to all the URLs using a "Fuzzy Membership Function" based on their user sessions support count. After assigning the weights to URLs and User Sessions, we apply a "Fuzzy c-Mean Clustering" algorithm [11] to discover the clusters of user profiles.

Fuzzy clustering techniques perform non-unique partitioning of the data items where each data point is assigned a membership value for each of the clusters. This allows the clusters to grow into their natural shapes [12]. A membership value of zero indicates that the data point is not a member of that cluster. A non-zero membership value shows the degree to which the data point represents a cluster. Fuzzy clustering algorithms can handle the outliers by assigning them very small membership degree for the surrounding clusters. Thus fuzzy clustering is more robust method for handling natural data with vagueness and uncertainty.

Rest of the paper is organized as follows: in section-II, we provide a detailed description of our fuzzy set theoretic approach for feature evaluation and dimensionality reduction by assigning weights to URL items and user sessions. In section III, we discuss how to apply Fuzzy c-Mean Clustering algorithm to the weighted as well as non-weighted sessionized data. Section IV gives the results of applying the Fuzzy c-Mean Clustering methodology to web log data generated by a real Web site, results are presented for clustering weighted as well as non-weighted sessionized data. Finally section V provides a brief conclusion.

II. FEATURE SELECTION AND SESSION WEIGHT ASSIGNMENT

We map the user sessions as vectors of URL references in a n-dimensional space. Let $U = \{u_1, u_2, \ldots, u_n\}$ be a set of n unique URLs appearing in the preprocessed log and let $S = \{s_1, s_2, \ldots, s_m\}$ be a set of m user sessions discovered by preprocessing the web log data, where each user session $s_i \in S$ can be represented as $s = \{w_{u_1}, w_{u_2}, \ldots, w_{u_m}\}$. Each $w_{u_i}$ may be either a binary or non-binary value depending on whether it represents presence and absence of the URL in the session or some other feature of the URL. If $w_{u_i}$ represents presence of absence of the URL in the session, then each user session is represented as a bit vector where

$$w_{u_i} = \begin{cases} 1; & \text{if } u_i \in s; \\ 0; & \text{otherwise} \end{cases} \quad (1)$$

Instead of binary weights, feature weights can also be used to represent a user session. These feature weights may be based on frequency of occurrence of a URL reference within the user session, the time a user spends on a particular page or the number of bytes downloaded by the user from a page.

Fuzzy Feature Evaluation and Dimensionality Reduction by Assigning Weights to the URL Items:

The URLs appearing in the access logs could number in the thousands. Distance-based clustering methods often perform very poor when dealing with very high dimensional data. Therefore filtering the logs by removing references to low support URLs (i.e. that are not supported by a specified number of user sessions) can provide an effective dimensionality reduction method while improving clustering results. We propose a "Fuzzy Set Theoretic" approach for filtering the low support URLs. We assign weights to all the URLs using a "Fuzzy Membership Function" based on their user sessions support count. In this approach we use lower and upper bounds for the user session support. All the URLs with session support count lower than the lower bound α1 are assigned the weight 0. On the all the URLs with session support count higher than the upper bound α2 are assigned the weight 1. All the URLs with session support count between α1 and α2 are assigned the weight between 0 and 1 using a linear fuzzy membership function.

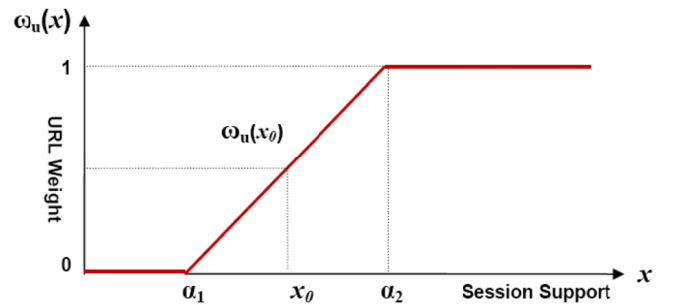

Fig. 1. Linear Fuzzy Membership Function for URL Weight Assignment

Fig. 1 depicts a linear fuzzy membership function for URL weight assignment. The URL weight assignment using the linear fuzzy membership function is carried out using the following formula:

$$\omega_u(x) = \begin{cases} 0, & \text{if } x \le \alpha_1 \\ 1, & \text{if } x \ge \alpha_2 \\ \dfrac{x - \alpha_1}{\alpha_2 - \alpha_1}, & \text{otherwise} \end{cases} \quad (2)$$

where

$\omega_u$ is weight assigned to the URL $u$

$\alpha_1$ is the lower threshold on the session support count

$\alpha_2$ is the upper threshold on the session support count

$x$ is the session support count of URL $u$

*A. Fuzzy Approach for Assigning Weights to User Sessions:*

Before performing the clustering of the user sessions, the session files can be filtered to remove very small sessions in order to eliminate the noise from the data [5]. But direct removal of these small sized sessions may result in loss of a significant amount of information especially when the number of small sessions is large. We propose a "Fuzzy Set Theoretic" approach to deal with this problem. Instead of directly removing all the small sessions below a specified threshold, we assign weights to all the sessions using a "Fuzzy Membership Function" based on the number of URLs accessed by the sessions.

Fig. 2 depicts a linear fuzzy membership function for session weight assignment. Here β₁ represents a lower threshold on the number of URLs accessed in a session and β₂ represents an upper threshold on the number of URLs accessed in a session.

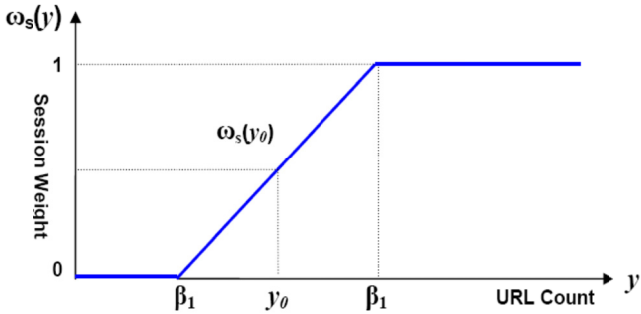

Fig. 2. Linear Fuzzy Membership Function for Session Weight Assignment

The session weight assignment using the linear fuzzy membership function takes is carried out using the following formula:

$$\omega_s(y) = \begin{cases} 0, & \text{if } |s| \le \beta_1 \\ 1, & \text{if } |s| \ge \beta_2 \\ \dfrac{|s| - \beta_1}{\beta_2 - \beta_1}, & \text{otherwise} \end{cases} \quad (3)$$

where $\omega_s$ is weight assigned to the session $s$

$\beta_1$ is the lower threshold on the session URL count

$\beta_2$ is the upper threshold on the session URL count

$y = |s|$ is the number of URLs accessed in session $s$

## III. Fuzzy c-means Clustering technique

Clustering techniques are widely used in WUM to capture similar interests and trends among users accessing a Web site. Clustering aims to divide a data set into groups or clusters where inter-cluster similarities are minimized while the intra cluster similarities are maximized. Details of various clustering techniques can be found in survey articles [13]-[15]. The ultimate goal of clustering is to assign data points to a finite system of k clusters. Union of these clusters is equal to a full dataset with the possible exception of outliers. Clustering groups the data objects based only on the information found in the data which describes the data objects and the relationships between them.

Fuzzy C-means clustering (FCM) incorporates the basic idea of Hard C-means clustering (HCM). In order to understand the Fuzzy C-means clustering let's first discuss the Hard c-means clustering algorithm, and then we will explore the details of Fuzzy c-means clustering algorithm.

*A. Hard c-Means Clustering :*

The Hard *c*-means clustering technique is [16] is a popular technique and has been applied to a variety of areas [17]-[19]. This algorithm relies on finding cluster centres by trying to minimize a cost function of dissimilarity measure. The Hard *c*-Means clustering algorithm [16] is one of the most commonly used methods for partitioning the data. Given a set of *m* data points $X = \{x_i \mid i = 1 \cdots m\}$, where each data point is a *n*-dimensional vector, *k*-means clustering algorithm aims to partition the *m* data points into *k* clusters ($k \le m$) $C = \{c_1, c_2, \ldots, c_k\}$ so as to minimize an objective function $J(V, X)$ of dissimilarity, which is the within-cluster sum of squares. In most cases the dissimilarity measure is chosen as the Euclidean distance. The objective function is an indicator of the distance of the *n* data points from their respective cluster centers. The objective function $J$, based on the distance between a data point $x_i$ in cluster $j$ and the corresponding cluster centre $v_j$, is defined in (4).

$$J(X,V) = \sum_{j=1}^{k} J_i(x_i, v_j) = \sum_{j=1}^{k}\left(\sum_{i=1}^{m} u_{ij}.d^2(x_i, v_j)\right), \quad (4)$$

where, $J_i(x_i, v_j) = \sum_{i=1}^{m} u_{ij}.d^2(x_i, v_j)$,

is the objective function within cluster $c_i$,

$u_{ij} = 1$, if $x_i \in c_j$ and 0 otherwise.

$d^2(x_i, v_j)$ is the disatnce between $x_i$ and $v_j$

Euclidian distance between various data points and cluster centers can be calculated using (5).

$$d^2(x_i, v_j) = \left\| \sum_{k=1}^{n} x_k^i - v_k^j \right\|^2 \quad (5)$$

where, $n$ is the number of dimensions of each data point

$x_k^i$ is the value of $k^{th}$ dimensions of $x_i$

$v_k^j$ is the value of $k^{th}$ dimensions of $v_j$

The partitioned clusters are defined by a $m \times k$ binary membership matrix $U$, where the element $u_{ij}$ is 1, if the *i*th data point $x_i$ belongs to the cluster $j$, and 0 otherwise. Once the cluster centers $V = \{v_1, v_2, ..., v_k\}$, are fixed, the membership function $u_{ij}$ that minimizes (4) can be derived as follows:

$$u_{ij} = \begin{cases} 1; & \text{if } d^2(x_i, v_j) \leq d^2(x_i, v_{j^*}) \, j \neq j^*, \forall j^* = 1, \cdots, k \\ 0; & \text{otherwise} \end{cases} \quad (6)$$

The equation (5) specifies that assign each data point $x_i$ to the cluster $c_j$ with the closest cluster center $v_j$. Once the membership matrix $U=[u_{ij}]$ is fixed, the optimal center $v_j$ that minimizes (4) is the mean of all the data point vectors in cluster $j$:

$$v_j = \frac{1}{|c_j|} \sum_{i, x_i \in c_j}^{m} x_i \quad (7)$$

where,

$|c_j|$, is the size of cluster $c_j$ and also $|c_j| = \sum_{i=1}^{m} u_{ij}$

### B. Fuzzy c-Means Clustering:

Fuzzy c-means clustering was proposed by Dunn [20] and improved by Bezdek [21] and is frequently used in the field of data mining. In this technique each data point belongs to a cluster to a degree specified by a membership grade. As in hard c-means clustering, Fuzzy C-means clustering relies on minimizing a cost function of dissimilarity measure. Fuzzy C-means clustering (FCM) incorporates the basic idea of Hard C-means clustering (HCM), with the difference that in FCM each data point belongs to a cluster to a degree of membership grade, while in HCM every data point either belongs to a certain cluster or does not belong. Therefore FCM performs the fuzzy clustering in such a way that a given data point may belong to several clusters with the degree of belongingness specified by membership grades between 0 and 1. However, FCM still uses an objective function that is to be minimized while trying to partition the data set. The algorithm calculates the cluster centers and assigns a membership value to each data item corresponding to every cluster within a range of 0 to 1. The algorithm utilizes a fuzziness index parameter q where $q \in [1, \infty]$ [22] which determines the degree of fuzziness in the clusters. As the value of q reaches to 1, the algorithm works like a crisp partitioning algorithm. Increase in the value of q results in more overlapping of the clusters.

Let $X = \{x_i \mid i = 1 \cdots m\}$ be a set of $n$-dimensional data point vectors where $m$ is the number of data points and each $x_i = \{x_1^i, x_2^i, \cdots, x_n^i\} \forall i = 1 \cdots m$. Let $V = \{x_j \mid j = 1 \cdots c\}$ represent a set of n-dimensional vectors corresponding to the cluster center corresponding to each of the c clusters and each $v_j = \{v_1^j, v_2^j, \cdots, v_n^j\} \forall j = 1 \cdots c$ Let uij represent the grade of membership of data point xi in cluster j. $u_{ij} \in [1, 0] \, \forall i = 1 \cdots m$ and $\forall j = 1 \cdots c$. The $n \times c$ matrix $U = [u_{ij}]$ is a fuzzy c-partition matrix, which describes the allocation of the data points to various clusters and satisfies the following conditions:

$$\left. \begin{array}{l} \sum_{j=1}^{c} u_{ij} = 1, \, \forall i = 1 \cdots m \\ 0 < \sum_{j=1}^{c} u_{ij} < m, \, \forall j = 1 \cdots c \end{array} \right\} \quad (8)$$

The performance index $J(U,V,X)$ of fuzzy c-mean clustering can be specified as the weighted sum of distances between the data points and the corresponding centers of the clusters. In general it takes on the form:

$$J(U, V, X) = \sum_{j=1}^{c} \sum_{i=1}^{m} u_{ij}^{q} d_{ij}^{2} \left( \overline{x}_i, \overline{v}_j \right) \quad (9)$$

where,

$q \in [1, \infty]$ is the fuzziness index of the clustering

$d_{ij}^2 \left( \overline{x}_i, \overline{v}_j \right)$ is the distance between $\overline{x}_i$ and $\overline{v}_j$

The distance $d_{ij}^2 \left( \overline{x}_i, \overline{v}_j \right)$ for weighted and non-weighted user sessions and URLs is calculated using equations (10) and (11) respectively:

$$d_{ij}^{2} \left( \overline{x}_i, \overline{v}_j \right) = \sum_{k=1}^{n} w_s(x_i) \left\| w_u \left( x_k^i \right) . \overline{x}_k^{i} - \overline{v}_k^{j} \right\| \quad (10)$$

$$d_{ij}^{2} \left( \overline{x}_i, \overline{v}_j \right) = \sum_{k=1}^{n} \left\| \overline{x}_k^{i} - \overline{v}_k^{j} \right\| \quad (11)$$

$w_s(x_i)$ is the weight of the data point $x_i$ and

$w_u \left( x_k^i \right)$ is the weight of $k^{th}$ feature the data point $x_i$

Minimization of the performance Index $J(U,V,X)$ is usually achieved by updating the grade of memberships of data points and centers of the clusters in an alternating fashion until convergence. This performance Index is based on the sum of the squares criterion. During each of the iterations, the cluster centers are updated as follows:

$$\overline{v}_j = \frac{\sum_{i=1}^{m} u_{ij}^q \overline{x}_i}{\sum_{i=1}^{m} u_{ij}^q} \quad (12)$$

Membership values are calculated by the following formula:

$$u_{ij} = \frac{\left( \frac{1}{d_{ij}^2 \left( \overline{x}_i, \overline{v}_j \right)} \right)^{1/(q-1)}}{\sum_{k=1}^{n} \left( \frac{1}{d_{ij}^2 \left( \overline{x}_i, \overline{v}_j \right)} \right)^{1/(q-1)}} \quad (13)$$

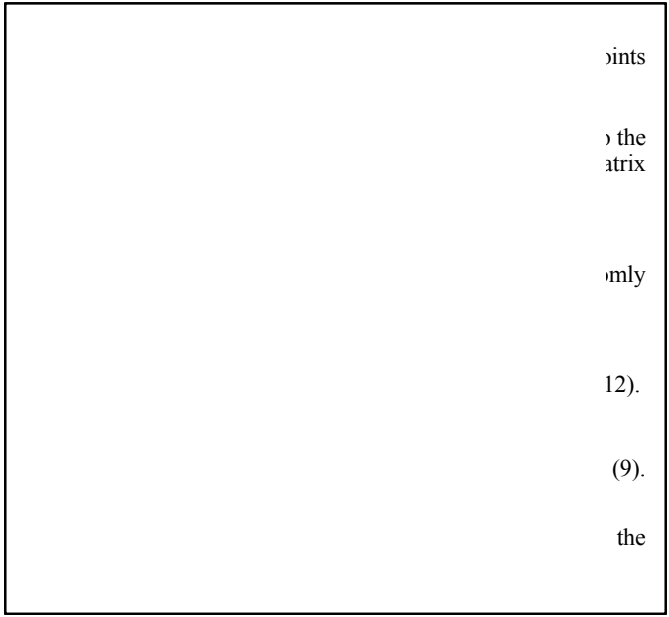

Fig. 3. Fuzzy c-Means Clustering Algorithm

Fig. 3 describes the algorithm for Fuzzy C-Means Clustering. In order to decide the number of optimum clusters for the data set *X* we use a validity function *S* which is the ratio of compactness to separation [22] as given below:

$$S = \frac{\sum_{j=1}^{c}\sum_{i=1}^{m} u_{ij}^2 \left\| \bar{x}_i - \bar{v}_j \right\|^2}{m . \min_{l \neq k} \left\| \bar{v}_l - \bar{v}_k \right\|^2} \quad (14)$$

for each $c = c_{\min}, \cdots, c_{\max}$

Let $\Omega_c$ denote the optimal candidate at each *c* then, the solution to the following minimization problem yields the most valid fuzzy clustering of the data set.

$$\min_{c_{\min} \leq c \leq c_{\max}} \left( \min_{\Omega_c} S \right) \quad (15)$$

IV. EXPERIMENTAL RESULTS

The Web access logs are taken from the P.A. College of Engineering, web site, at URL http://www.pace.edu.in. The site hosts a variety of information, including departments, faculty members, research areas, and course information. The Web access logs covered a period of one month, from February 1, 2011 to February 8, 2011. There were 12744 logged requests in total. In order to discover the clusters that exist in user access sessions of a web site, following steps are performed:
1. Pre-processed the web log data to discover the user sessions.
2. Performed the fuzzy feature evaluations and dimensionality reduction by assigning weights to the user sessions and URLs.
3. Applied the Fuzzy c-means clustering algorithm to the non-weighted user sessions and URLs obtained in step 1.
4. Applied the Fuzzy c-means clustering algorithm to the weighted user sessions and URLs obtained in step 2.
5. Compared the results of steps 4 and 5.

Details of the above experimental steps are described in the following sub-sections.

*A. Web Usage Log Preprocessing*

After performing the cleaning operation the output file contained 11995 entries. Total no. of unique users identified is 16 and the no. of user sessions discovered are 206. Table II depicts the results of cleaning and user session identification steps. Further details pre-processing can be found from our previous work [23].

TABLE I
RESULTS OF CLEANING AND USER IDENTIFICATION

| Items | Count |
|---|---|
| Initial No of Log Entries | 12744 |
| Log Entries after Cleaning | 11995 |
| No. of site ULRs accessed | 260 |
| No of Users Identified | 16 |
| No. of User Sessions Identified | 418 |

*B. Fuzzy Feature Evaluation and Dimensionality Reduction*

Table I shows that the number of URLs appearing in the access log is 260. Since each user session is represented as a vector of URL items, we would like to reduce the dimensionality of user session vectors by evaluating the URL items and eliminating the most insignificant ones. We evaluate each URL item based on the session support count. For this purpose we choose the lower bound on the number of sessions supported by the URL (LB) as 1 and an upper bound on the number of sessions supported by the URL (UB) as 6. Using equation (2), weights assigned to various URL items are described in Table I.

TABLE II
URL WEIGHTS BASED ON THE URL ACCESS COUNT

| URL Session Support | 1 | 2 | 3 | 4 | 5 | 6+ |
|---|---|---|---|---|---|---|
| URL Weight | 0 | 0.2 | 0.4 | 0.6 | 0.8 | 1 |

Fig. 4 shows the result of URL evaluation. Note that 142 URLs have the session support count of 1, hence they are eliminated by assigning them weight 0. The elimination of 142 URL items is a big reduction in the dimensionality of each user session vector. Other less significant URLs are assigned weights as show in Table II.

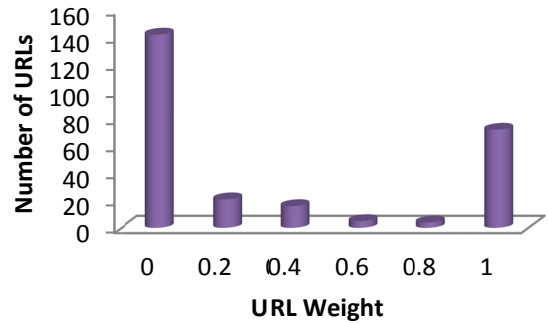

Fig. 4. Number of URLs vs. Associated URL Weight

## C. Assigning Weights to User Sessions:

Table I shows that the number of user sessions discovered from the access log are 418. Since each user session is represented by a row in the data matrix, we would like to reduce the row dimensionality of the data matrix by evaluating the user sessions and eliminating the most insignificant ones. We evaluate each user session based on the number of URL items accessed in that session. For this purpose we choose the lower bound on the number of URLs accessed in a session (LB) as 1 and an upper bound on the number of URLs accessed in a session (UB) as 6. Using equation (3), weights assigned to various sessions are described in Table II.

TABLE III
SESSION WEIGHTS BASED ON THE URL ACCESS COUNT

| Session URL Count | 1 | 2 | 3 | 4 | 5 | 6+ |
|---|---|---|---|---|---|---|
| Session Weight | 0 | 0.2 | 0.4 | 0.6 | 0.8 | 1 |

Fig. 5 shows the result of user session weight assignment. Note that 212 use sessions access only a single URL, hence they are eliminated by assigning them a weight of 0. The elimination of 212 user sessions is a big reduction in the row dimensionality of the data matrix. Other less significant user sessions are assigned weights as show in Table III.

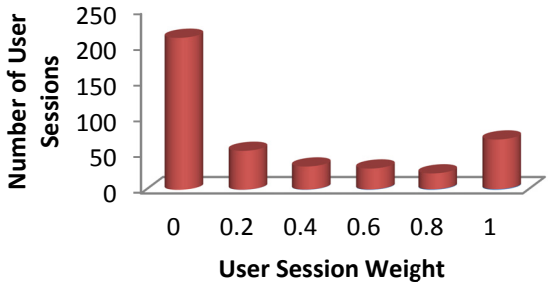

Fig. 5. Number of User Sessions vs. Associated Session Weight

The result of dimensionality reduction is show in the Table III below:

TABLE IV
RESULTS AFTER DIMENSIONALITY REDUCTION

| Items | Before Dimensionality Reduction | After Dimensionality Reduction |
|---|---|---|
| No. of URL items | 260 | 118 |
| No. of User Sessions | 418 | 206 |

## D. Mining User Session Clusters using Fuzzy c-Means Clustering :

Fuzzy c-Mean clustering algorithm (Fig. 3) is applied to discover session clusters that represent similar URL access patterns. The performance Index $J(U,V,X)$ of Fuzzy c-mean clustering is calculated using equation (9). It is the weighted sum of distances between the data points and the corresponding centers of the clusters.

Fuzzy c-Mean clustering algorithms described in Fig. 3 is first applied by initializing $k=2$. During each of the iterations we increased the number of clusters by 1 till the number of clusters is reached to 67 (One third of total number of user sessions). We repeated the above process for weighted as well as non-weighted URLs and sessions. While applying the c-means clustering algorithm to weighted and non-weighted user sessions, the main difference lies in calculating the distance between data points and cluster centres. For weighted sessions and URLs these distances are computed using equation (10). On the other hand for non-weighted user sessions, distances are calculated using equation (11).

Graph is Fig. 6 shows the performance index ($J$) versus number of clusters for weighted as well as non-weighted URLs and sessions. From the graph it is clear that "Fuzzy Set Theoretic" weight assignment to sessions and URLs results in better minimization of the performance index than non-weighted session approach.

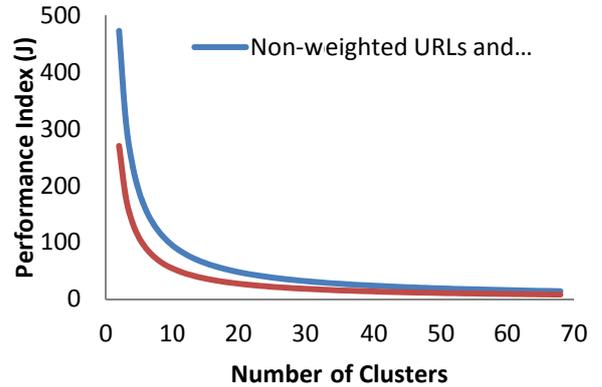

Fig. 6. Performance Index $J$ vs. No. Of Clusters $k$

In order to decide the number of optimum clusters we calculated the validity index ($S$), which is the ratio of compactness to separation using the equation (14).

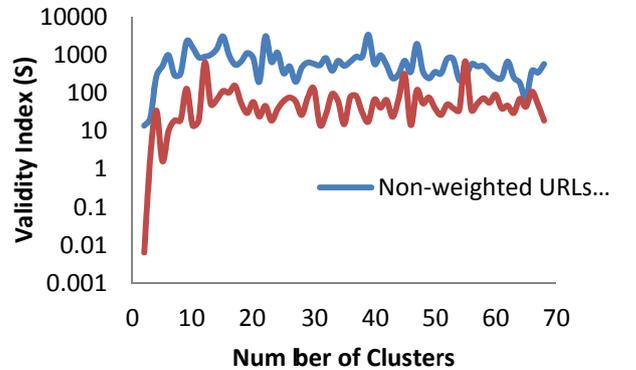

Fig. 7. Validity Index $S$ vs. No. Of Clusters $k$

Fig. 7 provides the graphs of validity index ($S$) versus number of clusters for weighted and non-weighted URLs and user sessions respectively. Our results show that the fuzzy approach of URL and session weight assignment resulted in better minimization of clustering validity index than without session weight assignment. Our result also shows that the validity index is minimized when $k=2$, for weighted as well as non-weighted URLs and sessions.

## V. CONCLUSION

In this paper, we discussed our methodology to perform feature evaluation and dimensionality reduction by assigning weights to URLs and user sessions using linear fuzzy membership functions. We also discussed the mathematical details about how to apply the Fuzzy c- Mean Clustering algorithm in order to cluster the user sessions.

In order improve the quality of the clustering; we proposed a "Fuzzy Set Theoretic" approach for feature evaluation and dimensionality reduction. Instead of directly removing all the low session support URLs below a specified threshold, we assign weights to the URLs using a fuzzy membership function based on the session support count of the URL. Similarly, instead of directly removing all the small sessions below a specified threshold, we assign weights to the sessions using another fuzzy Membership function based on the number of URLs accessed by the sessions. Finally we compared our soft computing based approach of URL and session weight assignment with the traditional hard computing based approach of low support URL and small session elimination. Our results show that the fuzzy set theoretic approach of URL and session weight assignment results in better minimization of clustering performance index than without session weight assignment. It also improves the validity index much better than without weight assignment.

## REFERENCES

[1] P. Kolari and A. Joshi, "Web mining: research and practice," Computing in Science and Engineering, vol. 6, no. 4, pp. 49–53, 2004.

[2] J. Srivastava, R. Cooley, M. Deshpande, and P.N. Tan. Web usage mining: Discovery and applications of usage patterns from web data. SIGKDD explorations, 1(2):12–23, 2000.

[3] W. Tong and H. Pi-lian, "Web log mining by an improved aprioriall algorithm," in In proceeding of world academy of science, engineering, and technology, 2005, pp. 97–100.

[4] R. W. Cooley, "Web usage mining: Discovery and application of interesting patterns from web data," Ph.D. dissertation, The Graduate School of the University of Minnesota, 2000.

[5] B. Mobasher. Data mining for web personalization. Lecture Notes in Computer Science, 4321:90, 2007.

[6] R. Cooley, B. Mobasher, J. Srivastava et al., "Data preparation for mining world wide web browsing patterns," Knowledge and Information Systems, vol. 1, no. 1, pp. 5–32, 1999.

[7] D. Tanasa and B. Trousse, "Advanced data preprocessing for intersites web usage mining," IEEE Intelligent Systems, vol. 19, no. 2, pp. 59–65, 2004.

[8] D. Tanasa and B. Trousse, "Data preprocessing for wum," Intelligent Systems, IEEE, vol. 23, no. 3, pp. 22–25, 2004.

[9] M. Spiliopoulou, B. Mobasher, B. Berendt, and M. Nakagawa, "A framework for the evaluation of session reconstruction heuristics in webusage analysis," INFORMS J. on Computing, vol. 15, pp. 171–190, April 2003.

[10] B. Mobasher, R. Cooley, and J. Srivastava, "Automatic personalization based on web usage mining," Commun. ACM, vol. 43, pp. 142–151, August 2000.

[11] James C. Bezdek, Robert Ehrlich and William Full, FCM: The Fuzzy c-Means Clustering Algorithm, Computers & Geosciences Volume 10, Issues 2-3, Pages 191-203, 1984.

[12] F. Klawonn and A. Keller, "Fuzzy clustering based on modified distance measures," in Advances in Intelligent Data Analysis, ser. Lecture Notes in Computer Science, D. Hand, J. Kok, and M. Berthold, Eds. Springer Berlin / Heidelberg, 1999, vol. 1642, pp. 291–301.

[13] P. Berkhin, "Survey of clustering data mining techniques," Springer, 2002.

[14] B. Pavel, "A survey of clustering data mining techniques," in Grouping Multidimensional Data. Springer Berlin Heidelberg, 2006, pp. 25–71.

[15] R. Xu and I. Wunsch, D., "Survey of clustering algorithms," Neural Networks, IEEE Transactions on, vol. 16, no. 3, pp. 645–678, May 2005.

[16] J. A. Hartigan and M. A. Wong, "A k-means clustering algorithm, Applied Statistics, 28:100--108, 1979.

[17] R.O. Duda, P.E. Hart, Pattern Classification and Scene Analysis, Wiley, New York, 1973.

[18] A.K. Jain, R.C. Bubes, Algorithm for Clustering Data, Prentice-Hall, Englewood Cli s, NJ, 1988.

[19] L. Kaufman, P.J. Rousseeuw, Finding Groups in Data. An Introduction to Cluster Analysis, Wiley, New York, 1990.

[20] J. C. Dunn, "A Fuzzy Relative of the ISODATA Process and Its Use in Detecting Compact Well-Separated Clusters", *Journal of Cybernetics* 3: 32-57, 1973

[21] J. C. Bezdek, "Pattern Recognition with Fuzzy Objective Function Algoritms", Plenum Press, New York. 1981

[22] X. L. Xie and G. Beni, "A validity measure for fuzzy clustering," IEEE Trans. Pattern Analysis and Machine Intelligence, vol. PAMI-13, p. 841847, 1987.

[23] Z. Ansari, M. F. Azeem, A. V. Babu, and W. Ahmed, "Preprocessing users web page navigational data to discover usage patterns," in The Seventh International Conference on Computing and Information Technology, Bangkok, Thailand, May 2011, proceeding vol. 1 pp. 18-189.